\pdfoutput=1
\documentclass[10pt,twocolumn,letterpaper]{article}

\usepackage[pagenumbers]{cvpr}

\usepackage{microtype}

\renewcommand{\paragraph}[1]{\vspace{.5em}\noindent\textbf{#1}\hspace{1em}}

\usepackage{verbatim}
\usepackage{xspace}
\newcommand{\dsname}{\mbox{EvoMusicFake-17K}\xspace}

\definecolor{cvprblue}{rgb}{0.21,0.49,0.74}
\usepackage[pagebackref,breaklinks,colorlinks,allcolors=cvprblue]{hyperref}

\title{IncrementDET: Continual AI-Generated Music Detection via Enrollment}

\author{Runbao Du\\
University of Chicago\\
{\tt\small runbao@uchicago.edu}
\and
Wei Li\textsuperscript{*}\\
Fudan University\\
{\tt\small weili-fudan@fudan.edu.cn}
\and
Qile He\textsuperscript{*}\\
Fudan University\\
{\tt\small 26113050261@m.fudan.edu.cn}
}

\begin{document}
\maketitle
{\renewcommand{\thefootnote}{}\footnotetext{\textsuperscript{*}Co-corresponding authors. Qile He is the first corresponding author.}}

\begin{abstract}
AI-generated music has become common on music streaming platforms, and listeners often cannot distinguish it from music made by humans. AI-music detection is therefore needed, but detectors trained on a fixed set of generators often fail on unseen generators. As new generators emerge frequently, updating a universal detector, a single model for all generators, typically requires substantial training data and computation, while risking performance degradation on previously learned generators. 
We therefore introduce IncrementDET, a detector that learns each new generator by a process called \emph{enrollment}. For each generator, IncrementDET trains one linear classifier, which we call a needle, to separate samples of that generator from a fixed set of human reference tracks. Enrolling a new generator adds one needle while the previously constructed needles remain unchanged. A \v{S}id\'ak correction keeps the overall false-positive rate at its target as needles are added.
To compare detectors under these conditions, we define the task of \emph{continual AI-generated music detection} and introduce a benchmark in which generators arrive one at a time. The benchmark measures the accuracy of a detector before and after each update and the cost of each update. For this benchmark, we also construct \dsname, a dataset of instrumental music from six recent generators and four human music platforms.
Compared with SpecTTTra, a state-of-the-art universal detector, IncrementDET learns a new generator more accurately from only a few samples, generalizes better to unseen generators and better preserves detection performance on previously learned generators, at a substantially lower update cost. The structure of IncrementDET can also be applied to the detection of generated images and video, where new generators are released just as frequently.
\end{abstract}
\begin{figure}[t]
\centering
\includegraphics[width=\linewidth]{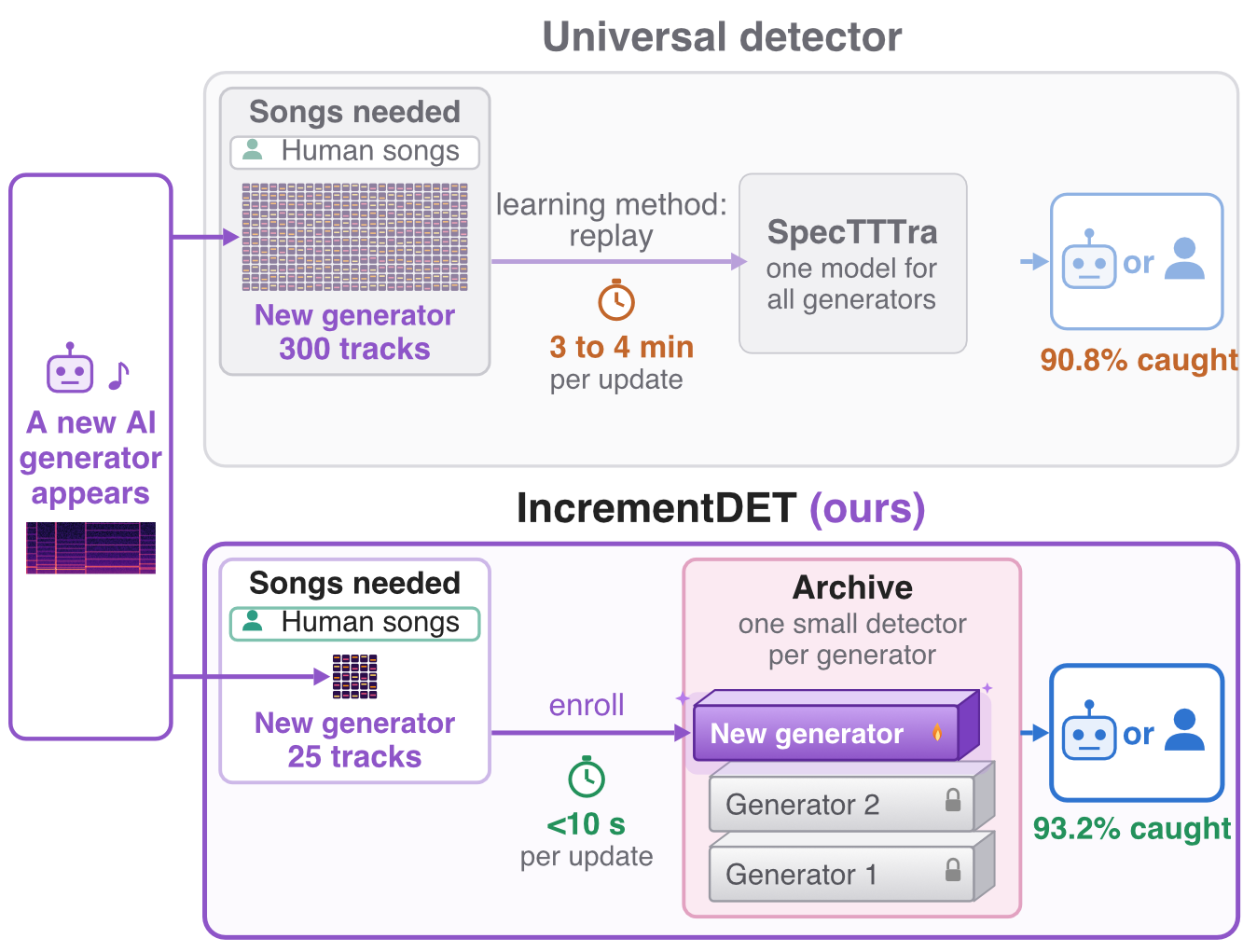}
\caption{\textbf{A schematic of our work.} Top: the universal detector SpecTTTra is updated by experience replay with $300$ tracks of the new generator. Bottom: IncrementDET passes only $25$ tracks of the new generator through a frozen encoder and adds one linear classifier (a needle), leaving earlier needles unchanged. Each row shows the time of one update and the TPR at $5\%$ FPR on the new generator, averaged over twelve generators (\cref{tab:fewshot}), so IncrementDET is more accurate with fewer tracks and far less compute.}
\label{fig:teaser}
\vspace{-8pt}
\end{figure}

\section{Introduction}
\label{sec:intro}

The rapid proliferation of AI-generated music on streaming platforms raises growing concerns about content authenticity and the protection of human artists. Deezer reported that fully AI-generated tracks exceeded half of its daily submissions on certain days in June 2026, and in a blind listening test with $9{,}000$ participants, $97\%$ could not distinguish fully AI-generated music from human-made music~\cite{deezer2026ai}. These developments underscore the need for reliable AI-generated music detection to ensure content transparency and support responsible platform governance.

However, maintaining a reliable universal detector remains challenging due to limited generalization to unseen generators (\cref{sec:related}) and the rapid emergence of new ones. Between 2023 and 2025, at least twelve text-to-music generators were released publicly~\cite{copet2023musicgen,liu2024audioldm2,chen2023musicldm,melechovsky2024mustango,evans2024stableaudioopen,zhang2025inspiremusic,ning2025diffrhythm,jiang2025diffrhythm2,gong2025acestep,levo2025,mureka,suno}, so a universal detector must be updated whenever a new generator appears, often when only a few samples of the new generator are available.

For music, however, no prior work has studied how detectors should adapt to emerging generators. Existing benchmarks such as SONICS~\cite{sonics2025}, FakeMusicCaps~\cite{comanducci2024fmc} and MoM~\cite{batra2025mom} evaluate a detector once on a fixed split of generators, and some of them hold out generators to test generalization, but none of them measures how well a detector learns a new generator. We therefore define a new task, \emph{continual AI-generated music detection}, and introduce a benchmark for it, together with a new dataset of generated and human music. In this benchmark, twelve generators arrive sequentially, and at each step every detector receives the \emph{same} samples of the new generator and the \emph{same} human tracks. We measure how accurately each detector identifies the new generator before and after its update, how much it forgets about earlier generators, and how much compute each update costs.

Existing detectors struggle to adapt to new generators efficiently without forgetting earlier ones. Larger models and more training data improve generalization, yet detection accuracy remains substantially lower on unseen generators (\cref{sec:related}). Human-only approaches, such as MusicDET~\cite{han2026musicdet}, model human music and flag low-likelihood tracks as AI-generated. However, density models may assign high likelihoods to out-of-distribution inputs~\cite{nalisnick2019know}, potentially missing new generators, while their human-only training paradigm offers no mechanism to adapt using newly available generated samples. Universal detectors can instead learn from new generators, but require substantial amounts of newly generated training data that may be difficult to obtain. Moreover, fine-tuning risks forgetting earlier generators~\cite{mccloskey1989catastrophic}, whereas full retraining requires retaining historical samples and incurs high computational costs. These limitations motivate a detector that learns new generators accurately from few samples, with minimal computational overhead and without compromising previously learned detection capabilities.

We thus propose IncrementDET, a detector that meets these requirements by learning each new generator through a cheap update, which we call \emph{enrollment}. This detector represents every track by the features of a frozen music encoder (MuQ), pooled over time. For each generator, it trains one logistic-regression classifier, which we call a \emph{needle}, to separate samples of that generator from a fixed set of human reference tracks. Each needle has its own threshold, which is set on a separate set of human calibration tracks, and a \v{S}id\'ak correction sets the false-positive rate of each needle, so that the false-positive rate of the detector as a whole stays at its target. A track is flagged as AI-generated if any needle scores it above its threshold. Enrolling a new generator therefore trains one needle on a small number of its samples in a few seconds and modifies no existing needle (\cref{fig:teaser}).

We compare IncrementDET with SpecTTTra~\cite{sonics2025}, a state-of-the-art universal detector that we update with several strategies, up to full retraining on all stored samples. IncrementDET outperforms it both in the standard cross-generator evaluation, with a lower EER on all twelve held-out generators, and on our continual benchmark, before and after it learns each new generator. Before a new generator is enrolled, IncrementDET detects it at a mean TPR of $57.2\%$ at $5\%$ FPR, against at most $27.4\%$ for SpecTTTra under any update strategy. At every step of the sequence, IncrementDET detects the generators learned so far more accurately than SpecTTTra under any update strategy. After the last step, it still detects the first generator at a mean TPR of $96.4\%$ at $5\%$ FPR, against $79.4\%$ for SpecTTTra with full retraining. With only $25$ tracks of a new generator, IncrementDET detects it at a mean TPR of $93.2\%$ and keeps $99.2\%$ on the other generators, whereas fine-tuned SpecTTTra reaches $84.6\%$ and falls to $43.1\%$. 

In summary, we make the following contributions.
\begin{itemize}
\item We define the task of continual AI-generated music detection and introduce a benchmark and \dsname, a dataset of generated and human instrumental music.
\item We propose IncrementDET, an enrollment-based detector that learns new generators through independent linear classifiers (\emph{needles}), with a \v{S}id\'ak correction to control the overall false-positive rate as generators accumulate.
\item Extensive experiments demonstrate that IncrementDET outperforms the state-of-the-art universal detector in both the standard cross-generator evaluation and our continual detection benchmark, achieving higher accuracy, better knowledge retention, and lower update costs.
\item We show that universal detectors struggle to adapt to new generators, requiring substantial training data and costly retraining to avoid forgetting earlier generators, whereas enrollment learns new generators from few samples at low computational cost while preserving prior performance.
\end{itemize}

\section{Related Work}
\label{sec:related}

\paragraph{Audio deepfake detection models.}
Most detectors of generated audio are discriminative classifiers trained to separate human-produced recordings from AI-generated ones. The first designs came from speech anti-spoofing. RawNet2~\cite{tak2021rawnet2} learns features directly from the raw waveform, and AASIST~\cite{jung2022aasist} applies graph attention to spectral and temporal features. Later detectors use a pretrained self-supervised encoder, such as wav2vec~2.0~\cite{baevski2020wav2vec2} or XLS-R~\cite{babu2022xlsr}, to extract features from the audio and train a small classifier, such as AASIST~\cite{tak2022wav2vec}, on these features. Many detectors of AI-generated music adopt these speech architectures directly, for example by combining the music encoder MERT~\cite{li2024mert} with an AASIST classifier~\cite{han2026musicdet}. Other detectors are designed specifically for music. SONICS~\cite{sonics2025} introduced SpecTTTra, a transformer that splits the spectrogram into separate spectral and temporal tokens so that it can process long songs efficiently. Further approaches use dual-stream contrastive learning~\cite{batra2025mom}, music-intrinsic features~\cite{sofia2026}, forensic residuals~\cite{artifactnet2026}, or classifiers on CLAP~\cite{wu2023clap} embeddings~\cite{crosvila2025}. In this paper, we use SpecTTTra as the universal detector, because it is a common baseline in AI-generated music detection~\cite{batra2025mom,han2026musicdet,crosvila2025} and its public training code allows us to update it at every step with the same data as IncrementDET.

\paragraph{Cross-generator generalization.}\label{sec:related_crossgen}
Previous studies report that detectors of AI-generated music perform poorly on unseen generators. Detectors trained on MusicGen and tested on Mustango reach EERs of $38.9\%$ for SpecTTTra-$\alpha$, $48.7\%$ for wav2vec2-AASIST and $46.4\%$ for MERT-AASIST~\cite{han2026musicdet}. On SONICS, every evaluated architecture detects the generators seen in training better than unseen ones, and the $F_1$ score of SpecTTTra-$\alpha$ falls to $0.56$ and $0.25$ on two later generators, SeedMusic and SkyMusic~\cite{sonics2025}. SpecTTTra trained on SONICS reaches an $F_1$ score of $0.972$ on SONICS but only $0.535$ on Riffusion and $0.688$ on YuE~\cite{batra2025mom}. A detector trained once on a fixed set of generators therefore cannot be expected to remain accurate as new generators are released.

\paragraph{Continual learning.}
Updating a detector as new generators arrive is a continual-learning problem, in which a model learns a sequence of tasks one at a time and must maintain its performance on the earlier ones. This is difficult because training only on new data causes catastrophic forgetting~\cite{mccloskey1989catastrophic,french1999catastrophic}. Experience replay~\cite{rolnick2019replay} and elastic weight consolidation~\cite{kirkpatrick2017ewc} reduce this forgetting, and methods designed for speech deepfake detectors, such as RAWM~\cite{zhang2023rawm} and RegO~\cite{chen2025rego}, reduce it by restricting how the weights of the detector change during each update. Continual detection has been benchmarked for image~\cite{li2023cddb} and speech deepfakes~\cite{zhang2024evda}, but not for music. E3~\cite{azizpour2024e3} trains one expert per new image generator from limited data, but fuses the experts with a network retrained on a memory buffer at every update. The design of IncrementDET is closest to class-incremental methods that keep a pretrained model frozen, such as SimpleCIL~\cite{zhou2024simplecil} and RanPAC~\cite{mcdonnell2023ranpac}, where each new class is represented by a prototype so that no earlier class is forgotten. These methods, however, classify among known classes without controlling a false-positive rate.

\paragraph{Enrollment.}
Our use of the term \emph{enrollment} comes from speaker verification, where a new speaker is added by averaging the embeddings of a few of their utterances, and a test utterance is accepted only if its embedding is close enough to this average~\cite{variani2014dvector}. Like SimpleCIL and RanPAC, IncrementDET adds each generator from a few samples without retraining, and like speaker verification, it makes a threshold decision for each generator. In addition, it trains a linear classifier against human music for each generator and calibrates its thresholds on human data. 
\section{Continual AI-Music Detection Benchmark}
\label{sec:setup}

\subsection{Evaluation protocol}
\label{sec:protocol}
In continual AI-generated music detection, generators $g_1,\dots,g_T$ arrive one at a time in a fixed order, and a detector must separate their outputs from human-made music, whose distribution we denote by $H$. A detection system is a sequence $D_0,D_1,\dots,D_T$, where $D_{t-1}$ is fixed before $g_t$ arrives and $D_t$ is the detector obtained after the system has used the samples of $g_t$ provided at its arrival. We measure two quantities:
\begin{align}
\textsc{Pre}(t) &= \text{performance of } D_{t-1} \text{ on } g_t, \\
\textsc{Post}(t) &= \text{performance of } D_t \text{ on } g_1,\dots,g_t \text{ and } H .
\end{align}
$D_0$ has seen no generator, so $\textsc{Pre}$ is reported for $t \ge 2$. $\textsc{Pre}$ measures what a detector achieves without an update, and $\textsc{Post}$ measures whether the update recovers detection of the new generator without reducing detection of earlier ones.

At each arrival, every detector receives the same $N$ labeled samples of $g_t$ and the same human tracks, and we record the GPU time of its update.

\subsection{\dsname dataset}
\label{sec:dataset}

We construct a dataset, \dsname, for this benchmark. It contains $6{,}000$ tracks ($115$\,h) generated by six recent generators, and $11{,}111$ human tracks ($651$\,h) collected from four platforms. In addition to \dsname, the benchmark uses the generated tracks of five generators from FakeMusicCaps~\cite{comanducci2024fmc}, Suno~\cite{suno} tracks that we generated but do not redistribute, and human tracks from the public datasets FMA~\cite{defferrard2017fma} and MagnaTagATune~\cite{law2009magnatagatune}. All tracks are instrumental, and \cref{sec:discussion} discusses this choice. Human tracks are longer than generated ones, but every detector scores only one $5$-second window per track. We will release the prompts, generation settings, track identifiers and collection scripts of \dsname.

\begin{figure*}[t]
\centering
\includegraphics[width=\linewidth]{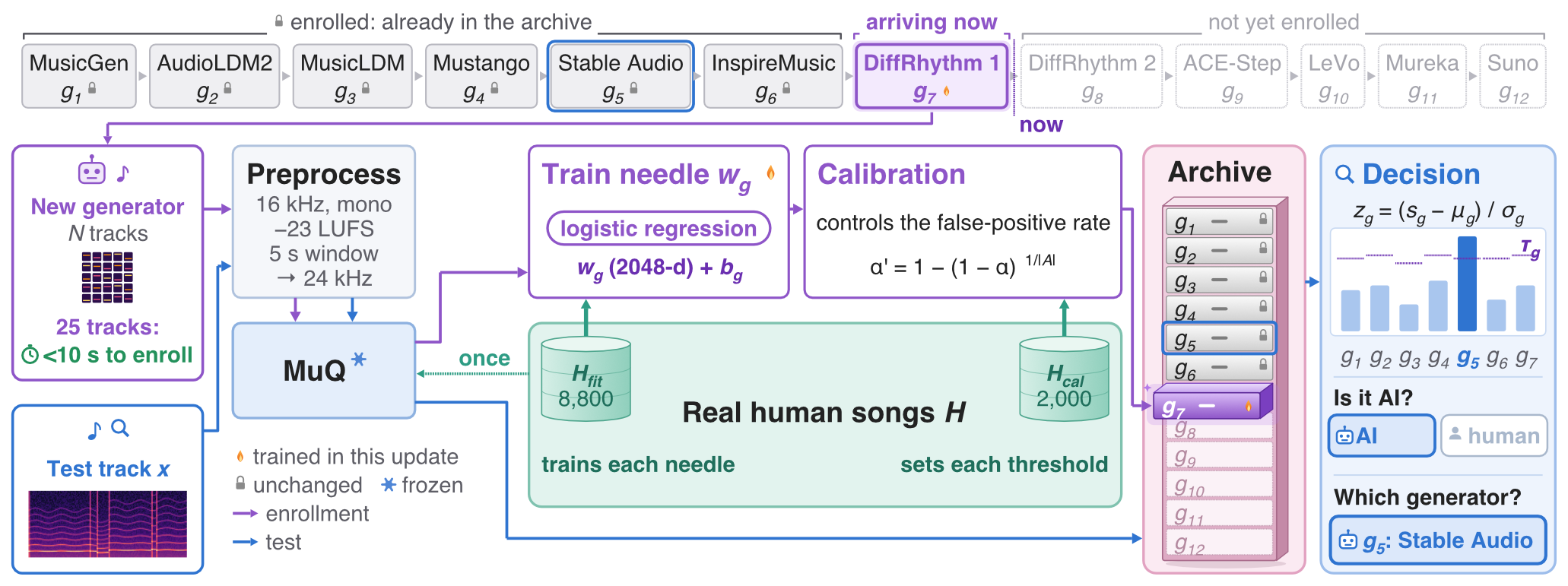}
\caption{\textbf{Overview of IncrementDET.} Top: the generators in their order of arrival, with DiffRhythm~1 arriving and the earlier generators already enrolled. Purple arrows show enrollment. The $N$ tracks of the new generator are preprocessed and encoded by the frozen MuQ encoder, a needle is trained against human reference tracks from $\mathcal{H}_{\text{fit}}$ and calibrated on $\mathcal{H}_{\text{cal}}$, and it is added to the archive without changing earlier needles. Blue arrows show a test. A test track is scored by every needle in the archive, and the decision panel flags it as AI-generated if any needle exceeds its threshold and names the generator whose needle gives the largest standardized score.}
\label{fig:method}
\vspace{-6pt}
\end{figure*}

\paragraph{AI generators.}\label{sec:data}
We use $T=12$ text-to-music generators whose first public versions were released between 2023 and 2025. Their families cover the main designs of current music generators (\cref{tab:generators}). The tracks of five generators are taken from FakeMusicCaps, which provides $10$-second clips generated from MusicCaps captions~\cite{agostinelli2023musiclm}. We generated the tracks of the other seven ourselves. The six generators in \dsname receive the same $1{,}000$ prompts for instrumental music, $125$ for each of eight genres. Suno receives $500$ of these prompts with two outputs each.

\begin{table}[t]
\centering
\footnotesize
\setlength{\tabcolsep}{4pt}
\caption{Generators, in the order in which they arrive. LM: language model. FM: flow matching. Transf.: transformer.}
\label{tab:generators}
\begin{tabular*}{\linewidth}{@{\extracolsep{\fill}}lll@{}}
\toprule
Generator & Version & Family \\
\midrule
MusicGen~\cite{copet2023musicgen}                 & medium    & Autoregressive LM \\
AudioLDM2~\cite{liu2024audioldm2}                 & music     & Latent diffusion  \\
MusicLDM~\cite{chen2023musicldm}                  & --        & Latent diffusion  \\
Mustango~\cite{melechovsky2024mustango}           & --        & Latent diffusion  \\
Stable Audio Open~\cite{evans2024stableaudioopen} & 1.0       & Diffusion transf. \\
InspireMusic~\cite{zhang2025inspiremusic}         & 1.5B-Long & Tokens + FM       \\
DiffRhythm~1~\cite{ning2025diffrhythm}            & v1.2      & Flow matching     \\
DiffRhythm~2~\cite{jiang2025diffrhythm2}          & --        & Flow matching     \\
ACE-Step~\cite{gong2025acestep,gong2026acestep15} & v1.5      & LM + rendering    \\
LeVo~\cite{levo2025}                              & large     & LM + rendering    \\
Mureka~\cite{mureka}                              & mureka-9  & Undisclosed       \\
Suno~\cite{suno}                                  & v5, v4.5+ & Undisclosed       \\
\bottomrule
\end{tabular*}
\vspace{-6pt}
\end{table}

\paragraph{Human music.}\label{sec:human}
We collected human tracks from four platforms, Jamendo~\cite{jamendo}, Pixabay~\cite{pixabay}, ccMixter~\cite{ccmixter} and the Internet Archive netlabels collection~\cite{archive_netlabels}, and selected instrumental tracks using platform tags, excluding vocal-tagged tracks on Pixabay. Because some platforms accept generated uploads, we screened the two largest sources for generated music. On Jamendo, $74\%$ of the tracks were released before 2023, and we excluded accounts whose upload behavior indicates generated music, such as many uploads within a short period. On Pixabay, we excluded tracks that a separately trained detector flags as likely generated. From FMA, we kept tracks with the genre label Instrumental, and from MagnaTagATune, clips with the crowd-sourced tag ``no vocals''. ccMixter and the Internet Archive netlabels collection form a set of held-out sources that is never used for training or calibration. Tracks shorter than $10$\,s are excluded. The supplementary material lists all sources and licenses.

\paragraph{Data splits.}\label{sec:splits}
All splits are disjoint and are assigned by a fixed hash of the track identifier. Each generator has $800$ fitting tracks and $200$ test tracks. The human tracks of Jamendo, FMA, Pixabay and MagnaTagATune are divided into a fitting set $\mathcal{H}_{\text{fit}}$ of $8{,}800$ tracks, a model-selection set of $2{,}200$ tracks, a calibration set $\mathcal{H}_{\text{cal}}$ of $2{,}000$ tracks and a test set $\mathcal{H}_{\text{test}}$ of $2{,}000$ tracks. The held-out set $\mathcal{H}_{\text{new}}$ consists of the $815$ tracks from ccMixter and the Internet Archive netlabels collection. Generated tracks have no calibration split, because thresholds are set on human music alone. The model-selection set is used only to select checkpoints and hyperparameters of SpecTTTra. The supplementary material lists the number of tracks of each source in each split.

\paragraph{Preprocessing and shortcut removal.}\label{sec:preprocessing}
All audio is cropped to at most the central $240$\,s, resampled to $16$\,kHz mono and loudness-normalized to $-23$\,LUFS, and all detectors score the same $5$-second window of each track. This reduces cues from how platforms store audio rather than from the music, such as sample rate, bitrate, codec, loudness and channel layout. Sample rate, bitrate or leading silence alone can suffice for detection on existing datasets~\cite{crosvila2025,borzi2022synthetic,muller2021silence,smeu2025shortcuts}. The $5$-second window also removes a duration cue (\cref{sec:shortcuts}).

\subsection{Evaluation metrics}
\label{sec:metrics}
We report the equal error rate (EER) and the true-positive rate (TPR) at false-positive rates (FPR) of $5\%$ and $1\%$. The low false-positive rates matter because a detector with a low EER can still miss most generated tracks once its threshold is set so that few human tracks are flagged. All detectors set their thresholds on the same human calibration set $\mathcal{H}_{\text{cal}}$. Their false-positive rates are then measured on $\mathcal{H}_{\text{test}}$, which comes from the same sources as $\mathcal{H}_{\text{cal}}$, and on $\mathcal{H}_{\text{new}}$, which comes from sources that no detector has seen and tests whether the target rate holds for new sources. We also report attribution accuracy, the share of detected generated tracks that are assigned to the correct generator. Forgetting is measured by the TPR on the first generator after each later step.
\section{Method}
\label{sec:method}

\Cref{fig:method} gives an overview of IncrementDET. Each enrolled generator has a needle, a linear classifier on frozen music features that separates the generator from human reference tracks (\cref{sec:features,sec:needles}), and each needle has a threshold set on human calibration tracks with a \v{S}id\'ak correction (\cref{sec:decision}). Enrolling a new generator adds one needle and updates the thresholds (\cref{sec:enroll}), and at test time a track is flagged if any needle exceeds its threshold (\cref{sec:inference}).

\subsection{Feature extraction}
\label{sec:features}
Let $f$ be a frozen self-supervised audio encoder and $\bar{f}(x)\in\mathbb{R}^d$ the pooled representation of a track $x$. In all
experiments $f$ is MuQ~\cite{zhu2025muq} (checkpoint \texttt{MuQ-large-msd-iter}). The $5$-second window of
Sec.~\ref{sec:preprocessing} is resampled to $24$\,kHz, the hidden states of one encoder layer are pooled over time, and their mean and standard deviation are concatenated, so that $d=2048$. 
Freezing the encoder keeps every needle in the same feature space, so the features of the human reference set are extracted only once.

The layer is chosen from the encoder's $13$ layers by a fixed rule. For each layer and each generator available for training, a needle is trained on $80\%$ of that generator's fitting tracks and evaluated on the remaining $20\%$, each against a separate set of human fitting tracks, and the layer with the lowest mean EER is used. In the leave-one-generator-out comparison, the rule is applied separately in each round, so the held-out generator never influences the choice. In the continual comparison, it is applied once, at the first step, with the fitting tracks of the first generator only. In every case the rule selects layer $0$ or layer $1$.

\subsection{Needles}
\label{sec:needles}
For each generator $g$ in the set $\mathcal{A}$ of enrolled generators, IncrementDET stores one needle $(w_g, b_g)$ with $w_g\in\mathbb{R}^d$. Given $N$ samples $\mathcal{X}_g$ of the generator, the needle is the solution of a regularized logistic regression,
\begin{equation}
\resizebox{0.9\linewidth}{!}{$\displaystyle
(w_g, b_g) = \arg\min_{w,\,b}\ \sum_{x \in \mathcal{X}_g \cup \mathcal{H}_{\text{ref}}}
\ell\!\left(y_x,\ \langle w, \bar{f}(x)\rangle + b\right) + \lambda\|w\|^2 ,$}
\end{equation}
with $y_x=1$ for generated and $0$ for human tracks, where $\mathcal{H}_{\text{ref}}$ is a fixed set of human tracks from $\mathcal{H}_{\text{fit}}$ (Sec.~\ref{sec:exp_setup}). Features are standardized on the needle's training set, $\ell$ is the class-balanced logistic loss, and $\lambda$ is fixed (inverse regularization strength $C=1$) for every generator.

We use a separate linear classifier for each generator instead of one shared classifier for three reasons. First, a needle is trained only on samples of its own generator and the human reference set, so enrolling a generator requires no stored samples of earlier generators and leaves their needles unchanged. Second, separate needles detect more generated tracks than one shared classifier at strict false-positive rates and with few samples of a new generator, and they keep the false-positive rate at its target (\cref{tab:heads}). Third, the needle that responds most strongly identifies the generator, so detection and attribution come from the same model. 

\subsection{Calibration and decision}
\label{sec:decision}
Each needle $g$ assigns a track $x$ the score $s_g(x)=\langle w_g,\bar{f}(x)\rangle + b_g$, and a track is labeled
AI-generated if the score of any needle exceeds that needle's threshold
$\tau_g$. Every enrolled needle therefore adds another chance of a false alarm on human music. We treat the decision over the growing set of needles as a multiple-comparison problem in which the number of tests grows with every enrollment. To our knowledge, no earlier AI-generated content detector controls this growth. If each of the $|\mathcal{A}|$ enrolled needles had a false-positive rate $\alpha'$ and their errors on human music were independent, the overall false-positive rate would be $1-(1-\alpha')^{|\mathcal{A}|}$. To keep it at $\alpha$, IncrementDET recomputes $\alpha'$ at every enrollment with the classical \v{S}id\'ak correction~\cite{sidak1967},
\begin{equation}
\alpha' = 1-(1-\alpha)^{1/|\mathcal{A}|},
\end{equation}
so that the overall false-positive rate is $\alpha$. The threshold $\tau_g$ of each needle is the $(1-\alpha')$ quantile of its scores on
$\mathcal{H}_{\text{cal}}$. With twelve needles and $\alpha=5\%$, $\alpha'\approx0.43\%$, so each threshold is set by about the nine highest-scoring of $2{,}000$ calibration tracks. The correction therefore requires a calibration set of thousands of tracks. The scores of different needles on human music are positively correlated, because all needles read the same features, so on human sources represented in the calibration set the correction is slightly conservative in practice, while on new sources the measured rate can exceed the target (Sec.~\ref{sec:ablations}).

The TPR at a fixed FPR uses this per-needle decision. The EER, which requires one score per track, uses the maximum standardized score $z^{\max}(x) = \max_{g \in \mathcal{A}} z_g(x)$, where $z_g(x) = (s_g(x) - \mu_g)/\sigma_g$ and $\mu_g$ and $\sigma_g$ are the mean and standard deviation of the needle's scores on $\mathcal{H}_{\text{cal}}$. Attribution returns $\arg\max_{g} z_g(x)$, the generator whose needle responds most strongly relative to its scores on human music.

\subsection{Enrollment}
\label{sec:enroll}
When a new generator $g$ arrives with $N$ tracks, IncrementDET performs four steps. It extracts the features of the $N$ tracks, trains the needle $(w_g,b_g)$, computes the scores of the needle on $\mathcal{H}_{\text{cal}}$ to obtain $\mu_g$ and $\sigma_g$, and recomputes $\alpha'$ and all thresholds from the stored calibration scores.

Enrollment has one cost. Because the \v{S}id\'ak correction tightens every threshold as more generators are enrolled, the true-positive rate of earlier needles can decrease slightly even though their directions are unchanged. This cost falls almost entirely on the weakest needles (Sec.~\ref{sec:ablations}).

\subsection{Inference and storage}
\label{sec:inference}
At test time, a track is passed through the encoder once, and all needles are applied at once. The standardization of each needle only shifts and rescales each feature, so it can be absorbed into the weights and the offset of the needle. Stacking the needles then gives all scores with one matrix-vector product,
\begin{equation}
\mathbf{s}(x) = W\bar{f}(x) + \mathbf{b},\qquad
W\in\mathbb{R}^{|\mathcal{A}|\times d},\ \mathbf{b}\in\mathbb{R}^{|\mathcal{A}|}.
\end{equation}
Scoring a track therefore costs one encoder pass and one dot product of length $d$ per needle. The encoder pass takes about $7.4$\,ms per window on a consumer GPU, whereas the dot products for twelve needles are negligible in comparison, so adding needles barely changes the inference time.

\section{Experiments}
\label{sec:experiments}

\subsection{Experimental setup}  
\label{sec:exp_setup}

\paragraph{Baselines.} We compare IncrementDET with SpecTTTra~\cite{sonics2025}, which serves as the universal detector. In the continual comparison, SpecTTTra is updated with five strategies, namely \emph{no update}, \emph{full retraining} on all generators seen so far, \emph{fine-tuning} on the new generator and the human tracks, \emph{experience replay}, which adds $50$ stored tracks of each earlier generator, and \emph{elastic weight consolidation} (EWC)~\cite{kirkpatrick2017ewc}. To test whether the advantage of IncrementDET comes from the MuQ encoder or from enrollment, we also update one logistic-regression head on the same frozen MuQ features by retraining, fine-tuning, or fine-tuning with replay, and a class-mean classifier as in SimpleCIL~\cite{zhou2024simplecil}.

\paragraph{Implementation details.} IncrementDET uses an overall false-positive target $\alpha$ of $5\%$ or $1\%$. SpecTTTra-$\alpha$ uses its official code and $5$-second configuration. Full retraining runs for $50$ epochs and the other strategies for $10$. All runs use one RTX~3060, and costs are wall-clock times excluding evaluation. The human reference set $\mathcal{H}_{\text{ref}}$ is all of $\mathcal{H}_{\text{fit}}$ in the leave-one-generator-out comparison and a fixed subset of $2{,}000$ of its tracks in the continual comparison, which uses $N=300$ fitting tracks per arriving generator. Thresholds are set on $\mathcal{H}_{\text{cal}}$, and $95\%$ confidence intervals come from $1{,}000$ bootstrap resamples of all test and calibration sets.

\subsection{Comparison with the universal detector}
\label{sec:comparison}

\subsubsection{Leave-one-generator-out comparison}
\label{sec:comp_static}
Each of the twelve generators is held out in turn, and both detectors are trained on the other eleven. IncrementDET detects the held-out generator with its eleven enrolled needles (Sec.~\ref{sec:decision}), so this experiment measures how well the enrolled needles transfer to a new generator.

\begin{table}[!ht]
\centering
\footnotesize
\setlength{\tabcolsep}{4pt}
\caption{Detection without updates (leave-one-generator-out). EER and TPR at $5\%$ and $1\%$ FPR, in percent.}
\label{tab:stage0}
\begin{tabular*}{\linewidth}{@{\extracolsep{\fill}}lcccccc@{}}
\toprule
& \multicolumn{3}{c}{SpecTTTra} & \multicolumn{3}{c}{IncrementDET} \\
\cmidrule(lr){2-4}\cmidrule(l){5-7}
Held out & EER & @5 & @1 & EER & @5 & @1 \\
\midrule
MusicGen          & 23.0 & 58.0 & 41.5 &  2.5 &  99.0 & 95.5 \\
AudioLDM2         & 12.0 & 80.0 & 63.5 &  1.1 &  99.0 & 99.0 \\
MusicLDM          & 35.5 & 15.5 &  1.5 &  6.5 &  90.0 & 66.0 \\
Mustango          & 24.0 & 42.0 & 22.5 &  0.5 & 100.0 & 99.5 \\
Stable Audio Open & 11.6 & 75.5 & 49.5 &  2.2 &  98.5 & 96.5 \\
InspireMusic      & 36.5 & 13.0 &  3.5 &  9.0 &  80.5 & 60.5 \\
DiffRhythm~1      & 47.6 &  4.5 &  1.0 & 31.9 &  14.5 &  6.5 \\
DiffRhythm~2      & 47.6 &  5.0 &  1.0 & 40.6 &   9.0 &  2.5 \\
ACE-Step          &  9.0 & 83.0 & 53.5 &  5.0 &  93.5 & 68.0 \\
LeVo              & 16.0 & 63.5 & 40.5 &  4.6 &  94.5 & 82.5 \\
Mureka            & 35.1 & 10.5 &  1.5 & 33.1 &  11.5 &  0.0 \\
Suno              & 30.5 & 34.5 & 11.0 & 21.9 &  25.5 &  9.5 \\
\midrule
Mean              & 27.4 & 40.4 & 24.2 & \textbf{13.2} & \textbf{68.0} & \textbf{57.2} \\
\bottomrule
\end{tabular*}
\vspace{-8pt}
\end{table}

\Cref{tab:stage0} reports all twelve rounds. The EER of IncrementDET is lower than that of SpecTTTra in all twelve rounds ($13.2\%$ against $27.4\%$ on average), significantly so in all but Mureka, and its TPR at $5\%$ FPR is higher in eleven, significantly so in nine. Four generators, DiffRhythm~1, DiffRhythm~2, Mureka and Suno, are difficult for both detectors. On them, the mean TPR at $5\%$ FPR is $15.1\%$ for IncrementDET and $13.6\%$ for SpecTTTra.

\paragraph{Transfer within architecture families.} A held-out generator is detected best when a generator of the same architecture family, a group of generators that produce audio in the same way (\cref{tab:generators}), is already enrolled. Seven of the twelve generators have a family member among the other eleven. For five of these seven, the needle of a family member gives the highest score on $60.6$--$98.5\%$ of the detected tracks. The exceptions are DiffRhythm~1 and DiffRhythm~2, two flow-matching models whose generation procedures differ. DiffRhythm~1 generates the full song jointly~\cite{ning2025diffrhythm}, whereas DiffRhythm~2 generates it in successive blocks~\cite{jiang2025diffrhythm2}. Generators without an enrolled family member can still be detected. Held-out MusicGen, Stable Audio Open and InspireMusic reach $2.2$--$9.0\%$ EER.

\begin{figure*}[t]
\centering
\includegraphics[width=\linewidth]{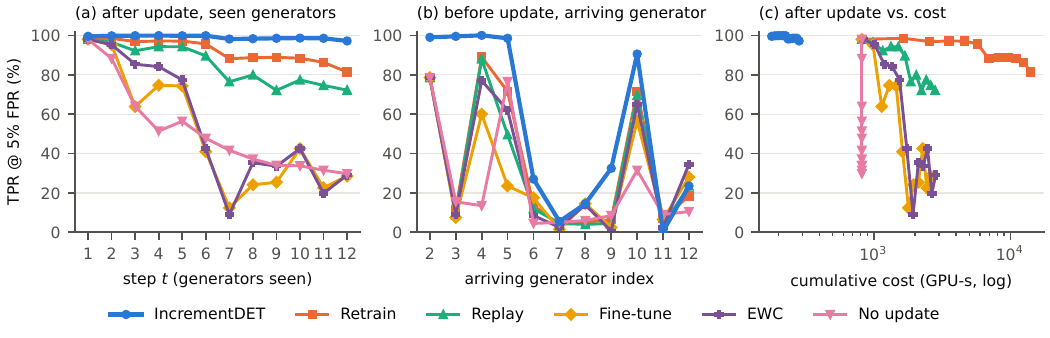}
\caption{Continual comparison in the order of \cref{tab:generators}, TPR at $5\%$ FPR. (a) After each update, mean over all generators seen so far. (b) Before each update, on the generator that is about to arrive. (c) Panel (a) against cumulative update cost.}
\label{fig:stage1}
\vspace{-6pt}
\end{figure*}

\paragraph{Geometry of needle directions.} The directions of the needles explain the family transfer of Sec.~\ref{sec:comp_static}. For MuQ, MERT~\cite{li2024mert}, wav2vec~2.0~\cite{baevski2020wav2vec2}, XLS-R~\cite{babu2022xlsr} and EnCodec~\cite{defossez2022encodec}, we standardize the features on the human fitting tracks, train one needle per generator and layer, and compute the cosine similarity between every pair of needles. In the four self-supervised encoders, needles of the same family are more similar than needles of different families, with a mean cosine similarity of $0.22$--$0.46$ within families against $0.10$--$0.20$ across them, and the difference exceeds that of at least $99.7\%$ of $5{,}000$ random relabelings that keep the family sizes. Across the $84$ pairs of a family and a generator outside it, higher similarity is rank-correlated with lower transfer EER, the EER on the outside generator of a detector fitted on all members of the family, with Spearman correlations of $-0.77$ to $-0.92$. For EnCodec, which has a single layer of codec features, the difference is not significant and the correlation is $-0.47$ (see the supplementary material). Enrolling one generator of a family therefore also improves the detection of its later members.

\subsubsection{Continual comparison}
\label{sec:comp_timeline}
The twelve generators arrive in the order of \cref{tab:generators} (\cref{fig:stage1}) and in four random orders (\cref{tab:timeline}).

\begin{table}[t]
\centering
\footnotesize
\setlength{\tabcolsep}{4pt}
\caption{Continual comparison, last step, mean ($\pm$ SD) over five orders. @5, @1: $\textsc{Post}(T)$ as TPR at $5\%$, $1\%$ FPR. Pre: mean of $\textsc{Pre}(t)$ over $t = 2,\dots,T$, as TPR at $5\%$ FPR. Stored: tracks of earlier generators kept for updates.}
\label{tab:timeline}
\begin{tabular*}{\linewidth}{@{\extracolsep{\fill}}lcccc@{}}
\toprule
Method & Stored & @5 & @1 & Pre \\
\midrule
\multicolumn{5}{@{}l}{\emph{SpecTTTra}} \\
\quad No update   & $0$    & $24.0 \pm 5.2$ & $14.5$ & $17.3$ \\
\quad Fine-tuning & $0$    & $30.0 \pm 3.0$ & $17.3$ & $17.8$ \\
\quad EWC         & $0$    & $37.8 \pm 6.4$ & $20.5$ & $20.1$ \\
\quad Replay      & $550$  & $74.5 \pm 2.1$ & $53.1$ & $24.8$ \\
\quad Retraining  & $3{,}300$ & $81.4 \pm 1.0$ & $56.0$ & $27.4$ \\
\midrule
IncrementDET      & $0$    & $\mathbf{97.2} \pm 0.0$ & $\mathbf{90.8}$ & $\mathbf{57.2}$ \\
\bottomrule
\end{tabular*}
\vspace{-6pt}
\end{table}

\paragraph{Detection and forgetting.} In the order of \cref{tab:generators} (\cref{fig:stage1}a), IncrementDET detects the generators seen so far at $97.2$--$99.9\%$ after every enrollment. It attributes $97.1$--$100\%$ of the detected tracks to the correct generator at every step. Full retraining is the best update strategy for SpecTTTra, but after twelve steps it detects only $81.4\%$ of the generated tracks, and replay, EWC and fine-tuning reach $74.5\%$, $37.8\%$ and $30.0\%$ (\cref{tab:timeline}). After the last step, IncrementDET still detects the first generator at $96.4\%$ on average over the five orders, against $79.4\%$ for full retraining. Fine-tuning learns each arriving generator but loses earlier ones, and EWC only delays this loss. After the last step in the order of \cref{tab:generators}, the EER of IncrementDET is at most $10.1\%$ on every generator, while that of retraining exceeds $11\%$ on five of the twelve. Before an update, no detector reaches the TPR that IncrementDET reaches after enrollment (\cref{fig:stage1}b).

\begin{table}[t]
\centering
\footnotesize
\setlength{\tabcolsep}{3pt}
\caption{IncrementDET and one classifier on the same frozen MuQ features. @5, @1, FPR: last step of the continual comparison, mean over five orders, where FPR is measured on $\mathcal{H}_{\text{test}}$ at the $5\%$ target. New, Old: TPR at $5\%$ FPR when a new generator arrives with $25$ tracks, on that generator and on the other eleven, mean over twelve generators. Stored: tracks of earlier generators kept for updates. Attr.: attributes a detected track to its generator.}
\label{tab:heads}
\begin{tabular*}{\linewidth}{@{\extracolsep{\fill}}lccccccc@{}}
\toprule
Method & Stored & @5 & @1 & FPR & New & Old & Attr. \\
\midrule
\multicolumn{8}{@{}l}{\emph{One classifier}} \\
\quad Retrained  & $3{,}300$ & $95.7$ & $77.7$ & $5.2$ & $82.3$ & $97.1$ & \texttimes \\
\quad Replay     & $550$  & $92.9$ & $73.9$ & $6.0$ & $80.0$ & $96.3$ & \texttimes \\
\quad Fine-tuned & $0$    & $84.4$ & $66.0$ & $5.6$ & $86.3$ & $96.0$ & \texttimes \\
Class means      & $0$    & $9.3$  & $4.8$  & $5.1$ & $5.0$  & $5.7$  & \checkmark \\
\midrule
IncrementDET     & $0$    & $\mathbf{97.2}$ & $\mathbf{90.8}$ & $\mathbf{3.5}$ & $\mathbf{93.2}$ & $\mathbf{99.2}$ & \checkmark \\
\bottomrule
\end{tabular*}
\vspace{-6pt}
\end{table}

\paragraph{Encoder or enrollment?} \Cref{tab:heads} compares IncrementDET with one classifier on the same frozen MuQ features. IncrementDET outperforms this classifier under all three update strategies. Even when the classifier is retrained at every step on all $3{,}300$ stored tracks, IncrementDET detects more generated tracks at $1\%$ FPR ($90.8\%$ against $77.7\%$, $95\%$ CI of the difference $[6.0, 19.8]$), and its false-positive rate stays below the $5\%$ target, whereas that of every classifier exceeds it. When a new generator arrives with only $25$ tracks, IncrementDET detects it at $93.2\%$, against $80.0$--$86.3\%$ for the classifier, and keeps $99.2\%$ on the other generators. It also attributes each detected track to its generator, which one classifier cannot do, and it stores no tracks of earlier generators. A classifier on class means, as in SimpleCIL, reaches only $9.3\%$, and RAWM~\cite{zhang2023rawm} and RegO~\cite{chen2025rego} do no better than fine-tuning (see the supplementary material).

\paragraph{Cost.} One enrollment of IncrementDET takes $9.2$--$13.8$\,s, about $9$\,s to extract the features of the $300$ new tracks and under one second to train the needle, plus a one-time cost of $165$\,s for the human reference set. One update of SpecTTTra takes $838$--$1{,}585$\,s under retraining and $157$--$201$\,s under the other strategies. Over the twelve steps, the cumulative cost of IncrementDET is $284$\,s, compared with $14{,}118$\,s for
retraining, about fifty times more (\cref{fig:stage1}c).

\subsection{Ablation and shortcut analysis}
\label{sec:ablations}\label{sec:shortcuts}

\paragraph{Number of enrollment samples.} Starting from each round of the leave-one-generator-out comparison, the held-out generator arrives with $N \in \{25,50,100,300,700\}$ tracks, where each smaller set is contained in every larger one. IncrementDET enrolls it with one needle, and SpecTTTra is fine-tuned on the $N$ tracks and the $2{,}000$ human tracks for $10$ epochs, or updated by replay. All models are tested on the test set of the continual comparison.

\begin{table}[!ht]
\centering
\footnotesize
\setlength{\tabcolsep}{4pt}
\caption{Enrollment with $N$ samples. TPR at $5\%$ FPR in percent, mean over the twelve held-out generators. New: the held-out generator. Old: the other eleven generators. $N=0$ is the model before the update. For SpecTTTra, it comes from a separate inference run and differs slightly from \cref{tab:stage0}.}
\label{tab:fewshot}
\begin{tabular*}{\linewidth}{@{\extracolsep{\fill}}rcccccc@{}}
\toprule
& \multicolumn{2}{c}{IncrementDET} & \multicolumn{2}{c}{Fine-tune} & \multicolumn{2}{c}{Replay} \\
\cmidrule(lr){2-3}\cmidrule(lr){4-5}\cmidrule(l){6-7}
$N$ & New & Old & New & Old & New & Old \\
\midrule
0   & \textbf{68.0} & \textbf{99.4} & 40.6 & 94.8 & 40.6 & 94.8 \\
25  & \textbf{93.2} & \textbf{99.2} & 84.6 & 43.1 & 58.5 & 91.3 \\
50  & \textbf{96.7} & \textbf{99.2} & 89.7 & 45.6 & 71.4 & 92.0 \\
100 & \textbf{97.3} & \textbf{99.2} & 93.7 & 48.4 & 81.4 & 91.0 \\
300 & \textbf{98.4} & \textbf{99.2} & 96.2 & 50.6 & 90.8 & 90.7 \\
700 & \textbf{98.8} & \textbf{99.2} & 97.2 & 47.5 & 94.1 & 89.5 \\
\bottomrule
\end{tabular*}
\vspace{-6pt}
\end{table}

As shown in \cref{tab:fewshot}, IncrementDET detects the new generator at $93.2\%$ with $25$ tracks and at $98.8\%$ with $700$, and its detection of the other eleven generators stays at $99.2\%$ for every $N$. Ten of the twelve generators reach at least $97.5\%$ with $25$ tracks; the exceptions are DiffRhythm~1 ($68.5\%$) and Suno ($57.0\%$). Fine-tuning SpecTTTra also detects the new generator well, but its detection of the other eleven generators falls
from $94.8\%$ to $43.1$--$50.6\%$ and does not recover as $N$ grows. Replay keeps the other generators at $89.5$--$92.0\%$ but needs $300$ tracks to reach $90.8\%$ on the new generator.

\paragraph{Effectiveness of the \v{S}id\'ak correction.} We repeat the continual comparison with each needle thresholded at the target rate, without the \v{S}id\'ak correction. At a $5\%$ target, the false-positive rate grows with the number of needles, from $3.8\%$ to $33.6\%$ on $\mathcal{H}_{\text{test}}$ and from $5.5\%$ to $41.0\%$ on $\mathcal{H}_{\text{new}}$. With the correction, it stays at $3.2$--$4.3\%$ on $\mathcal{H}_{\text{test}}$ and at $5.5$--$7.9\%$ on $\mathcal{H}_{\text{new}}$. The correction lowers the final mean TPR only from $99.8\%$ to $97.2\%$, and almost all of this loss falls on the two weakest needles, Suno ($99.5\%$ to $82.5\%$) and DiffRhythm~1 ($98.5\%$ to $85.5\%$), while the other generators lose at most $0.5$ points. Under MP3 encoding, resampling and equalization, IncrementDET stays ahead of every update strategy of SpecTTTra once thresholds are set on degraded human tracks (see the supplementary material).

\paragraph{Null needles.} To test whether a needle detects generation or merely a difference between sources, we treat each human source as a generator. The null needles of Jamendo, FMA, Pixabay, ccMixter and the Internet Archive netlabels reach EERs of $24.8$--$41.4\%$, against $0.6\%$ on average for the generator needles. Only MagnaTagATune is separated ($0.15\%$ EER), by its low-bitrate MP3 encoding, so the human calibration set should contain the expected encodings.

\paragraph{Duration shortcut.} The five FakeMusicCaps generators produce $10$-second clips, whereas all other sources contain longer tracks. With $120$-second inputs, track duration alone gives $0\%$ EER in all five held-out rounds of these generators, and SpecTTTra labels $100\%$ of human tracks as generated once they are truncated to $10$\,s and zero-padded, against $0.5\%$ for the same tracks unmodified. All detectors therefore score one unpadded $5$-second window of each track.

\section{Discussion and Limitations}
\label{sec:discussion}

\paragraph{Instrumental music.}
We restrict the dataset to instrumental music so that detection does not depend on the singing voice, on which detectors of singing-voice deepfakes rely almost entirely~\cite{chen2025singfake}. Detection on songs with vocals, which commercial generators such as Suno and Mureka produce by default, therefore remains to be tested. IncrementDET can be extended to such songs without changing its structure, by separating each track into vocals and accompaniment and enrolling a needle on each part, or by using an encoder that also represents the singing voice. The supplementary material discusses further limitations and potential harm.

\paragraph{Extension to vision.}
Only the encoder of IncrementDET is specific to audio, so our findings can inform the detection of generated images and video, where new generators are also released frequently. First, one linear classifier per generator on frozen features yields a more accurate detector at strict false-positive rates than one classifier retrained on all stored data (\cref{tab:heads}). Image detectors that use frozen features~\cite{ojha2023univfd} could therefore learn new generators by enrollment rather than by retraining. Second, when a detector adds one classifier per generator, its overall false-positive rate grows with every generator unless it is corrected (\cref{sec:ablations}). The \v{S}id\'ak correction that controls this rate does not depend on the modality. Third, the protocol of our benchmark evaluates a detector before and after each update and records the cost of each update, so it applies to other modalities.

\section{Conclusion}
\label{sec:conclusion}
We defined the task of continual AI-generated music detection, introduced a benchmark for it, constructed the \dsname dataset, and proposed IncrementDET, which enrolls each new generator as one needle. IncrementDET learns a new generator from only $25$ tracks, detects it at $93.2\%$ TPR at $5\%$ FPR, and keeps $99.2\%$ on the other generators. After twelve generators, it detects the generators seen so far at $97.2\%$ TPR, against $81.4\%$ for retraining the universal detector, with about one fiftieth of the update compute. A detector therefore does not need to be retrained to keep pace with new generators. Enrollment is more accurate, needs fewer samples and costs less, and its structure can be applied to other modalities.

\section*{LLM Usage Statement}
We used Claude (Anthropic), a large language model, throughout this work. In the writing, it revised passages in all sections, which the authors drafted and approved. In the research, it helped search the related literature, suggested some of the analyses, and reviewed much of the code for data processing, feature extraction, the experiments, the statistical analysis and the figures. The authors designed the study, decided which experiments to run, and verified all results, references and claims.

{
    \small
    \bibliographystyle{ieeenat_fullname}
    \bibliography{arxiv}
}

\twocolumn[\centering\Large\textbf{Supplementary Material}\vspace{1.5em}]
\appendix
\renewcommand{\thetable}{S\arabic{table}}
\renewcommand{\thefigure}{S\arabic{figure}}
\setcounter{table}{0}
\setcounter{figure}{0}

This supplementary material supports the claims of the main paper with further details and results. \Cref{sec:supp_data} documents the data and describes how we audited it to remove AI-generated tracks from the human data and to prevent shortcut learning from track duration. \Cref{sec:supp_heldout} shows that the advantage of IncrementDET on unseen generators holds for both data sources and does not depend on its configuration. \Cref{sec:supp_grouping} explains this advantage by the directions of the needles. \Cref{sec:supp_timeline} shows that IncrementDET leads in every generator order and against continual-learning methods for speech deepfake detection. \Cref{sec:supp_ablations} shows that the reported advantages are statistically significant and reports further ablations and the robustness to degradation. \Cref{sec:supp_limits} discusses limitations and future work.

\section{Data}
\label{sec:supp_data}
This section documents the sources and licenses of the data of Sec.~\ref{sec:dataset} and two audits of the data. The first removes AI-generated tracks from the human data, and the second prevents shortcut learning from track duration.

\subsection{Generators}
\Cref{tab:supp_gens} lists the version of each generator, the source of its tracks, the release date of that version and its license or terms of use. The tracks of MusicGen, AudioLDM2, MusicLDM, Mustango and Stable Audio Open are taken from FakeMusicCaps~\cite{comanducci2024fmc}, which is distributed under the Creative Commons Attribution-NonCommercial 4.0 license. We generated the tracks of the other seven generators ourselves. DiffRhythm~2 has no official instrumental mode, so its lyrics file contains only structure tags and no lyric lines, and its style prompt ends with ``instrumental, no vocals''. We listened to a sample of its outputs and found no vocals.

\begin{table*}[t]
\centering
\small
\caption{Generators in their order of arrival. Released: public release of the version that we use (year and month), which can be later than the first release of the generator. FMC: FakeMusicCaps. SAI CL: Stability AI Community License.}
\label{tab:supp_gens}
\begin{tabular}{@{}lllll@{}}
\toprule
Generator & Version & Source & Released & License or terms \\
\midrule
MusicGen~\cite{copet2023musicgen} & medium & FMC & 2023-06 & Code MIT, weights CC BY-NC 4.0 \\
AudioLDM2~\cite{liu2024audioldm2} & music checkpoint & FMC & 2023-08 & CC BY-NC-SA 4.0 \\
MusicLDM~\cite{chen2023musicldm} & -- & FMC & 2023-08 & CC BY-NC-SA 4.0 \\
Mustango~\cite{melechovsky2024mustango} & -- & FMC & 2023-11 & Code MIT, weights Apache 2.0 \\
Stable Audio Open~\cite{evans2024stableaudioopen} & 1.0 & FMC & 2024-06 & Stability AI Community License \\
InspireMusic~\cite{zhang2025inspiremusic} & 1.5B-Long & Ours & 2025-01 & Code Apache 2.0 \\
DiffRhythm~1~\cite{ning2025diffrhythm} & v1.2 & Ours & 2025-05 & Apache 2.0; VAE: SAI CL \\
DiffRhythm~2~\cite{jiang2025diffrhythm2} & -- & Ours & 2025-10 & Apache 2.0 \\
ACE-Step~\cite{gong2025acestep,gong2026acestep15} & v1.5 & Ours & 2026-01 & MIT \\
LeVo~\cite{levo2025} & SongGeneration-large & Ours & 2025-10 & Research and education use only \\
Mureka~\cite{mureka} & mureka-9 & Ours & 2026-04 & Mureka API terms of service \\
Suno~\cite{suno} & v5 and v4.5+ & Ours & 2025-09, 2025-07 & Suno Terms of Service \\
\bottomrule
\end{tabular}
\end{table*}

\subsection{Human music}
Jamendo~\cite{jamendo} tracks were obtained through the Jamendo API, and each track is distributed under the Creative Commons license chosen by its artist. FMA tracks come from the \texttt{fma\_large} subset of the FMA dataset~\cite{defferrard2017fma}, whose metadata is distributed under CC BY 4.0 and whose audio is distributed under the license chosen by each artist. MagnaTagATune clips~\cite{law2009magnatagatune} come from Magnatune and are distributed under the Creative Commons Attribution-NonCommercial-ShareAlike 3.0 license. Pixabay~\cite{pixabay} tracks are distributed under the Pixabay Content License. The $\mathcal{H}_{\text{new}}$ tracks come from ccMixter~\cite{ccmixter} and the Internet Archive netlabels collection~\cite{archive_netlabels}, and we use only items that carry a Creative Commons license. \Cref{tab:splits} lists the number of tracks of each source.

\begin{table}[h]
\centering
\small
\setlength{\tabcolsep}{2pt}
\caption{Data splits. Generated: tracks per generator. Human: tracks per source. Jam.: Jamendo; Pix.: Pixabay; MTAT: MagnaTagATune. The held-out set $\mathcal{H}_{\text{new}}$ comes from ccMixter ($365$) and the Internet Archive netlabels collection ($450$).}
\label{tab:splits}
\begin{tabular*}{\linewidth}{@{\extracolsep{\fill}}lcccccc@{}}
\toprule
& Generated & \multicolumn{5}{c}{Human} \\
\cmidrule(l){3-7}
Split & per gen. & Jam. & FMA & Pix. & MTAT & Total \\
\midrule
Fitting ($\mathcal{H}_{\text{fit}}$)       & 800 & 4,125 & 2,440 & 2,000 & 235 & 8,800 \\
Model selection                            & --  & 1,159 &  686 &  296 &  59 & 2,200 \\
Calibration ($\mathcal{H}_{\text{cal}}$)   & --  & 1,053 &  623 &  270 &  54 & 2,000 \\
Test ($\mathcal{H}_{\text{test}}$)         & 200 &  933 &  553 &  460 &  54 & 2,000 \\
Held-out ($\mathcal{H}_{\text{new}}$)      & --  & \multicolumn{4}{c}{other sources} & 815 \\
\bottomrule
\end{tabular*}
\end{table}

Some platforms accept generated uploads, so a human source can contain AI-generated tracks, and such tracks raise the thresholds that are set on human music. In an early audit, our per-generator needles flagged a set of outlying Jamendo tracks. The release dates and templated titles of the tracks, the registration dates of the uploading accounts, and the other uploads of the same accounts identified a group of uploading accounts whose tracks were probably generated. Removing $32$ of these tracks at the account level raised the TPR of the Suno needle at $5\%$ FPR from $47.3\%$ to $84.9\%$ in that audit. The effect is concentrated in the upper tail of the human score distribution, which is where strict thresholds are set, so it is not visible in the EER alone. The final data excludes $53$ tracks identified in this audit, in addition to the screening described in the main paper. Screening the human data is therefore necessary for detection at low false-positive rates.

\subsection{Duration Bias and Fixed-Length Evaluation}
\label{sec:supp_duration}
Differences in track-duration distributions between AI-generated and human music may introduce a \emph{duration-based shortcut}, allowing detectors to distinguish the two classes without learning meaningful acoustic characteristics. This bias can be particularly pronounced with long input windows, where shorter tracks are padded with silence, indirectly revealing their original durations.

To quantify this effect, we train a duration-only classifier using $\min(L,120\,\mathrm{s})$, where $L$ denotes the original track duration. In leave-one-generator-out evaluation, this classifier achieves $0\%$ EER on all five FakeMusicCaps rounds and at most $7.3\%$ EER on eight of the twelve rounds, despite having no access to audio content. These results demonstrate that duration alone can provide a strong discriminative cue, potentially inflating detection performance.

To eliminate this shortcut, we evaluate all detectors on fixed $5$-second windows without padding (\cref{sec:shortcuts}). Following the official SpecTTTra protocol~\cite{sonics2025}, the test window starts at $\lfloor 3(L-5\,\mathrm{s})/4 \rfloor$. This setting does not disadvantage SpecTTTra: when Suno is held out, its EER improves from $38.1\%$ with $120$-second inputs to $30.5\%$ with $5$-second inputs. Thus, fixed-length evaluation removes the duration cue while enabling a fairer comparison of acoustic detection capabilities.

\subsection{Use of data and human participants}
We use all audio only to compute features and to train and evaluate detectors for non-commercial research. We redistribute only audio that we generated ourselves, where the terms of the generator permit it, and no human audio. No human participants took part in this study, and we do not use artist names or other personal metadata as model input.

\section{Further results on unseen generators}
\label{sec:supp_heldout}
This section extends the leave-one-generator-out comparison of Sec.~\ref{sec:comp_static}. 

\subsection{Results on the two data sources}
IncrementDET has a lower mean EER than SpecTTTra both on the generators taken from FakeMusicCaps and on the generators whose tracks we generated ourselves. On the five FakeMusicCaps rounds, the mean EER is $2.6\%$ for IncrementDET and $21.2\%$ for SpecTTTra. On the other seven rounds, it is $20.9\%$ and $31.8\%$. The advantage of IncrementDET therefore holds for both groups and is not limited to the generators of FakeMusicCaps.

\subsection{Amount of training data}
The detection of unseen generators also needs little training data. With all fitting tracks, IncrementDET reaches $13.2\%$ EER and $68.0\%$ TPR at $5\%$ FPR on average over the twelve rounds. With only $50$ tracks per generator and $2{,}000$ human tracks, the needles still reach $14.1\%$ EER and $65.8\%$ TPR.

\subsection{Needle configuration}
We vary the configuration of IncrementDET in two ways and find that the simplest configuration of the needles is sufficient. Training each needle on five windows per track instead of one gives $14.3\%$ EER and $67.4\%$ TPR, and it raises the false-positive rate on the held-out human sources from $5.6\%$ to $8.6\%$. Adding a detector of atypical human music, which flags tracks that lie far from the human fitting tracks after PCA whitening, gives $14.6$--$14.9\%$ EER for $32$ to $256$ dimensions. 

\subsection{False positives on new human sources}
Both IncrementDET and SpecTTTra exceed the $5\%$ false-positive target on $\mathcal{H}_{\text{new}}$. However, IncrementDET exceeds it less, with an average false-positive rate of $5.6\%$ against $7.4\%$ for SpecTTTra. On $\mathcal{H}_{\text{test}}$, the false-positive rate of IncrementDET is $4.0$--$4.7\%$ in every round, so the target holds for the sources of $\mathcal{H}_{\text{cal}}$. The excess on $\mathcal{H}_{\text{new}}$ comes from one of its two sources. The false-positive rate of IncrementDET is $4.5\%$ on ccMixter and $6.6\%$ on the Internet Archive netlabels, and the models of SpecTTTra in the continual comparison show the same pattern. The excess is therefore caused by a source that differs from $\mathcal{H}_{\text{cal}}$, which means that a platform should calibrate its detector on tracks from its own catalog.

\section{Transfer within generator families}
\label{sec:supp_grouping}
This section gives the measurements behind the paragraph on the geometry of needle directions in Sec.~\ref{sec:comp_static}. Needles of generators from the same architecture family point in similar directions, and this similarity predicts how well a detector transfers to a generator that it has not seen. The finding gives an empirical explanation for the cross-generator generalization of IncrementDET and indicates its potential, because enrolling one generator of a family also prepares the detector for later members of that family.

\begin{table}[h]
\centering
\small
\setlength{\tabcolsep}{5pt}
\caption{Geometry of needle directions: cosine similarity within and across families (mean over layers), and Spearman $\rho$ between similarity and transfer EER.}
\label{tab:geometry}
\begin{tabular*}{\linewidth}{@{\extracolsep{\fill}}lcccc@{}}
\toprule
Encoder & Within & Across & Gap & Spearman $\rho$ \\
\midrule
MuQ~\cite{zhu2025muq}       & 0.460 & 0.202 & $+0.26$ & $-0.92$ \\
MERT~\cite{li2024mert}      & 0.294 & 0.109 & $+0.19$ & $-0.88$ \\
wav2vec~2.0~\cite{baevski2020wav2vec2} & 0.220 & 0.098 & $+0.12$ & $-0.77$ \\
XLS-R~\cite{babu2022xlsr}     & 0.243 & 0.117 & $+0.13$ & $-0.92$ \\
EnCodec~\cite{defossez2022encodec}   & 0.123 & 0.079 & $+0.04$ & $-0.47$ \\
\bottomrule
\end{tabular*}
\end{table}

\subsection{Similarity of needle directions}
Similar needle directions predict transfer in four self-supervised encoders. For each of five encoders, \cref{tab:geometry} reports the cosine similarity between needle directions within and across families, and the Spearman correlation between this similarity and the transfer EER over $84$ pairs of a family and a generator outside it. In the four self-supervised encoders, the similarity within families exceeds the similarity across families by $0.12$--$0.26$, and higher similarity goes with lower transfer EER ($\rho$ from $-0.77$ to $-0.92$). This gap exceeds that of at least $99.7\%$ of $5{,}000$ random relabelings of the generators that keep the family sizes. For EnCodec, a codec with a single layer of features, the gap is $0.04$ and not significant, and $\rho$ is $-0.47$. The finding is therefore not specific to MuQ and holds for the self-supervised encoders.

\subsection{Clustering of the generators}
We also cluster the twelve generators by their needle directions alone, without their family labels. For each encoder, we average the cosine similarity over layers, form eight groups, one per family, by average linkage on one minus this similarity, and compare the groups with the families by the adjusted Rand index (ARI). The groups agree closely with the families for MERT and wav2vec~2.0 (ARI $0.78$, $p<0.001$), more weakly for MuQ ($0.33$, $p=0.043$), and not significantly for XLS-R and EnCodec. Here $p$ is the fraction of the $5{,}000$ random relabelings that give an ARI at least as large. Some groups recur across encoders. ACE-Step and LeVo fall in the same group in MERT, wav2vec~2.0 and EnCodec. AudioLDM2, MusicLDM and Mustango form one group in MERT and wav2vec~2.0 and fall in the same larger group in MuQ. Suno and Mureka are each other's nearest direction in MuQ, wav2vec~2.0 and XLS-R, and Suno is the nearest direction of Mureka in all five encoders. DiffRhythm~1 and DiffRhythm~2 are never in the same group, which agrees with their weak transfer in the leave-one-generator-out comparison.
\section{Further results on the continual benchmark}
\label{sec:supp_timeline}
This section extends the continual comparison of Sec.~\ref{sec:comp_timeline}. 

\subsection{Generator orders}
IncrementDET leads in every generator order. In the continual comparison, the twelve generators arrive one at a time, and the main paper uses the order of \cref{tab:generators}. Because the result of an update strategy can depend on which generators come first, we repeat the continual comparison with four further orders of arrival. Each of these orders is a random shuffle of the twelve generators, drawn with a fixed random seed so that it can be reproduced. Every order uses the same fitting, replay and human tracks, so only the order of arrival changes. \Cref{tab:timeline} of the main paper reports the mean over these five orders, and \cref{tab:supp_orders} gives the TPR after the last step in each order. In every order, IncrementDET has the highest TPR, followed by retraining and by replay. The result of IncrementDET is the same in all orders, because each needle is trained independently of the other needles. The ranking of fine-tuning, RegO, EWC and no update varies between the orders. 

\begin{table}[h]
\centering
\small
\setlength{\tabcolsep}{3.5pt}
\caption{TPR at $5\%$ FPR on all twelve generators after the last step, per generator order. R: the order of \cref{tab:generators}. 1--4: random orders.}
\label{tab:supp_orders}
\begin{tabular}{@{}lcccccc@{}}
\toprule
Detector & R & 1 & 2 & 3 & 4 & Mean \\
\midrule
SpecTTTra, no update  & 29.8 & 25.6 & 14.8 & 22.5 & 27.5 & 24.0 \\
SpecTTTra, fine-tuning & 28.5 & 27.7 & 35.9 & 29.7 & 28.2 & 30.0 \\
SpecTTTra, RegO       & 29.5 & 27.3 & 31.6 & 34.0 & 26.0 & 29.7 \\
SpecTTTra, EWC        & 29.1 & 38.5 & 47.2 & 41.6 & 32.5 & 37.8 \\
SpecTTTra, replay     & 72.2 & 76.5 & 76.5 & 75.4 & 71.9 & 74.5 \\
SpecTTTra, retraining & 81.4 & 81.7 & 79.9 & 81.2 & 83.0 & 81.4 \\
IncrementDET          & \textbf{97.2} & \textbf{97.2} & \textbf{97.2} & \textbf{97.2} & \textbf{97.2} & \textbf{97.2} \\
\bottomrule
\end{tabular}
\end{table}

\subsection{Detection before the update}
No update strategy prepares SpecTTTra for the next generator. \textsc{Pre} measures how well a detector detects a generator that is about to arrive, before the detector has been updated with any of its tracks (Sec.~\ref{sec:protocol}). A high \textsc{Pre} therefore means that the detector already transfers to new generators. Averaged over the five orders, \textsc{Pre} is $57.2\%$ for IncrementDET, against $27.4\%$ for SpecTTTra under retraining, $24.8\%$ under replay, $20.1\%$ under EWC, $17.8\%$ under fine-tuning and $17.3\%$ without updates. Updating SpecTTTra on more generators therefore does not make it ready for the next one. \Cref{fig:stage1}b of the main paper shows \textsc{Pre} in the order of \cref{tab:generators}. This order is the most favorable of the five orders for SpecTTTra and the least favorable for IncrementDET. Even in this order, SpecTTTra reaches $34.2\%$ under retraining, $32.2\%$ under EWC, $31.5\%$ under replay, $27.0\%$ under fine-tuning and $23.5\%$ without updates, whereas IncrementDET reaches $53.7\%$, its lowest value over the five orders. 

\subsection{Comparison with continual-learning methods}
IncrementDET outperforms RAWM~\cite{zhang2023rawm} and RegO~\cite{chen2025rego}, two continual-learning methods designed for speech deepfake detection. Continual detection has not been studied for generated music (Sec.~\ref{sec:related}), so these methods are the closest earlier work for our task. Both methods only restrict how the weights of a detector change during each update, so they do not depend on speech and apply to a detector of generated music without change. After the last step of the continual comparison, RegO on the MuQ features reaches $83.9\%$ TPR at $5\%$ FPR and $64.3\%$ at $1\%$ FPR, and RAWM reaches $44.7\%$ and $31.8\%$, against $97.2\%$ and $90.8\%$ for IncrementDET (\cref{tab:supp_cl}). We apply both methods to a linear classifier on the MuQ features, and we also apply RegO to SpecTTTra. We normalize the projection matrix of RAWM by its spectral norm instead of its Frobenius norm, because with the Frobenius norm of the official code the projection shrinks by a factor of about $45$ at every step. Neither method improves on plain fine-tuning in our benchmark. RegO performs like fine-tuning on both detectors. RAWM keeps the first generator at $99\%$ or higher but learns new generators poorly, probably because its projection suppresses the directions spanned by the human tracks, which are the same at every step.

\begin{table}[h]
\centering
\small
\setlength{\tabcolsep}{3pt}
\caption{Continual-learning methods for speech deepfake detection, last step of the continual comparison, mean ($\pm$ SD) over the five orders. No method stores tracks of earlier generators.}
\label{tab:supp_cl}
\begin{tabular}{@{}llccc@{}}
\toprule
Detector & Update & @5 & @1 & Pre \\
\midrule
MuQ + linear classifier & Fine-tuning & $85.6 \pm 0.4$ & $67.4$ & $59.7$ \\
                      & RegO        & $83.9 \pm 1.0$ & $64.3$ & $58.6$ \\
                      & RAWM        & $44.7 \pm 3.5$ & $31.8$ & $40.4$ \\
SpecTTTra             & Fine-tuning & $30.0 \pm 3.0$ & $17.3$ & $17.8$ \\
                      & RegO        & $29.7 \pm 2.9$ & $17.1$ & $17.3$ \\
\midrule
IncrementDET          & --          & $\mathbf{97.2} \pm 0.0$ & $\mathbf{90.8}$ & $57.2$ \\
\bottomrule
\end{tabular}
\end{table}

\section{Further ablations and robustness}
\label{sec:supp_ablations}
This section first shows that the advantages of IncrementDET that the main paper reports are statistically significant, and then extends the ablations of \cref{sec:ablations}. 

\subsection{Statistical significance}
The advantages of IncrementDET are statistically significant. We estimate the uncertainty of the main comparisons with a paired bootstrap. Each resample draws, independently and with replacement, the $200$ test tracks of every generator, $\mathcal{H}_{\text{cal}}$, $\mathcal{H}_{\text{test}}$ and $\mathcal{H}_{\text{new}}$, and the thresholds are recomputed on the resampled $\mathcal{H}_{\text{cal}}$, so the intervals include the variation of the thresholds. All detectors are evaluated on the same resamples, so the difference between two detectors is measured on the same tracks. In the leave-one-generator-out comparison, the human tracks are the same in every round and are drawn once per resample. We use $1{,}000$ resamples and report the $2.5$th and $97.5$th percentiles as the $95\%$ confidence interval. A difference is significant when its interval excludes zero. \Cref{tab:supp_ci} lists the differences. The advantage of IncrementDET over SpecTTTra is significant in the continual comparison, in the leave-one-generator-out comparison and with $25$ enrollment tracks. In the leave-one-generator-out comparison, the mean EER of IncrementDET is lower by $14.1$ points ($95\%$ CI $[12.9, 15.3]$). Per round, the difference in EER is significant in every round except Mureka ($-2.0$, $[-5.6, 3.0]$), and the difference in TPR at $5\%$ FPR is significant in every round except DiffRhythm~2, Mureka and Suno. Against one retrained classifier on the MuQ features, the advantage at $1\%$ FPR is $13.0$ points and is also significant.

\begin{table}[h]
\centering
\small
\footnotesize
\setlength{\tabcolsep}{1.5pt}
\caption{Differences between IncrementDET and other detectors with $95\%$ bootstrap confidence intervals, in percentage points. Continual: last step of the order of \cref{tab:generators}. Held-out: mean over the twelve rounds of the leave-one-generator-out comparison. $N$: new generator enrolled with $N$ tracks, mean over the twelve rounds (\cref{tab:fewshot}). ST: SpecTTTra. MuQ: one linear classifier on the MuQ features (\cref{tab:heads}).}
\label{tab:supp_ci}
\begin{tabular}{@{}llcc@{}}
\toprule
Experiment & Against & TPR@5 & TPR@1 \\
\midrule
Continual & ST retraining & $+15.8$ $[13.2, 18.2]$ & $+33.3$ $[27.3, 42.3]$ \\
         & ST replay & $+25.0$ $[22.1, 28.4]$ & $+41.8$ $[34.4, 54.4]$ \\
         & MuQ retraining & $+1.5$ $[-0.2, 2.9]$   & $+13.0$ $[6.0, 19.8]$ \\
Held-out & ST & $+27.5$ $[24.6, 30.5]$ & $+33.0$ $[28.1, 37.5]$ \\
$N=25$  & Fine-tuning & $+8.7$ $[6.0, 11.1]$   & -- \\
$N=700$ & Fine-tuning & $+1.6$ $[-0.3, 2.4]$   & -- \\
\bottomrule
\end{tabular}
\end{table}

\subsection{False-positive control}
Each enrolled needle introduces another source of false positives, making overall FPR control essential. 
Without a correction, the false-positive rate on $\mathcal{H}_{\text{test}}$ grows from $3.8\%$ to $33.6\%$ over the twelve generators at the $5\%$ target (\cref{sec:ablations}). With the \v{S}id\'ak correction, the false-positive rate on $\mathcal{H}_{\text{test}}$ after the last step of the continual comparison is $3.45\%$ at the $5\%$ target and $0.95\%$ at the $1\%$ target, so the correction meets both targets on the sources of $\mathcal{H}_{\text{cal}}$. We also compare the correction with an alternative rule that sets one threshold on the maximum standardized score $z^{\max}(x)$ of Sec.~\ref{sec:decision}, at its $(1-\alpha)$ quantile on $\mathcal{H}_{\text{cal}}$. This single threshold also meets both targets ($4.5\%$ and $0.55\%$). 
We favor the \v{S}id\'ak correction because it achieves a higher TPR at the stricter $1\%$ target ($90.8\%$ vs. $88.3\%$), with comparable performance at $5\%$ ($97.2\%$ vs. $97.7\%$). Moreover, it supports modular enrollment: each needle's threshold can be updated using only its stored calibration scores and the total number of needles, without recomputing the joint score distribution.

\subsection{Diversity of the human reference data}
Both the fitting set and the calibration set of IncrementDET should mix several human sources. We train and calibrate the needles of the last step of the continual comparison on human tracks from a single source, Jamendo or FMA, instead of the mixed set (\cref{tab:humansrc}). The calibration set must mix sources for the detector to meet the false-positive target. When it comes from a single source, the false-positive rate on $\mathcal{H}_{\text{test}}$ exceeds the $5\%$ target ($6.9\%$ and $11.2\%$) and reaches $70.4$--$74.1\%$ on MagnaTagATune. With the mixed calibration set, the false-positive rate on $\mathcal{H}_{\text{test}}$ stays within the target ($4.3\%$ and $3.8\%$), and the mean TPR at $5\%$ FPR changes little ($97.1\%$ and $96.0\%$, against $97.2\%$). A mixed fitting set further protects sources that would otherwise be absent from it. On the $54$ MagnaTagATune test tracks, single-source fitting raises the false-positive rate from $0.0\%$ to $16.7\%$ and $24.1\%$. IncrementDET therefore uses human tracks from four sources in both sets.

\begin{table}[h]
\centering
\small
\setlength{\tabcolsep}{3pt}
\caption{Human data for training and calibrating the needles, last step of the continual comparison. TPR and EER are means over the twelve generators. FPR at the $5\%$ target on $\mathcal{H}_{\text{test}}$ (all sources and MagnaTagATune only) and on $\mathcal{H}_{\text{new}}$. All values in percent.}
\label{tab:humansrc}
\begin{tabular}{@{}llcccccc@{}}
\toprule
Fitting & Calibration & @5 & @1 & EER & FPR & MTAT & Held-out \\
\midrule
Mixed   & Mixed   & 97.2 & 90.8 & 1.9 & 3.5  & 0.0  & 6.9 \\
Jamendo & Mixed   & 97.1 & 89.9 & 2.5 & 4.3  & 16.7 & 5.8 \\
FMA     & Mixed   & 96.0 & 87.5 & 2.8 & 3.8  & 24.1 & 2.9 \\
Jamendo & Jamendo & 96.4 & 89.4 & 2.8 & 6.9  & 70.4 & 7.7 \\
FMA     & FMA     & 98.9 & 93.5 & 2.8 & 11.2 & 74.1 & 8.7 \\
\bottomrule
\end{tabular}
\end{table}

\subsection{Null needles}
To test whether a needle separates any two groups of tracks, we treat each human source as a generator (the null needles of \cref{sec:ablations}). A null needle is fitted on $300$ tracks of one source against the human negatives of the other sources, calibrated on $\mathcal{H}_{\text{cal}}$ without that source, and tested on held-out tracks of the source against the $\mathcal{H}_{\text{test}}$ tracks of the other sources. The null needles of Jamendo, FMA and Pixabay reach an EER of $41.4\%$, $36.0\%$ and $24.8\%$ (mean of five draws), and those of the two held-out sources, ccMixter and the Internet Archive netlabels ($200$ fitting tracks each), reach $37.6\%$ and $30.8\%$. Under the same protocol, the twelve generator needles reach at most $4.1\%$ EER. A needle therefore separates generated music from human music far better than it separates one human source from another. 

\subsection{Robustness to degradation}
Audio degradation does not remove the separation between generated and human music, but it invalidates thresholds that were set on clean audio. After recalibration on degraded human tracks, IncrementDET has a higher TPR than every update strategy of SpecTTTra under every degradation (\cref{tab:degrade}). We apply four degradations to every test window, generated and human. They are MP3 encoding at $128$\,kbit/s (and at $64$\,kbit/s), resampling to $11.025$\,kHz and back, a low-shelf filter at $200$\,Hz with $+6$\,dB gain, and a high-shelf filter at $4$\,kHz with $-6$\,dB gain. \Cref{tab:degrade_full} reports the EER and the false-positive rate with thresholds set on the clean calibration set. Equalization changes the EER of every detector by at most $1.5$ points. MP3 encoding and resampling raise the EER of IncrementDET from $1.9\%$ to $9.2\%$ and $14.3\%$, and that of retraining from $9.3\%$ to $21.5\%$ and $24.5\%$, so IncrementDET keeps the lowest EER under every degradation. They also shift the scores of human tracks, so that thresholds set on clean data no longer hold, and the false-positive rate rises to $25.6$--$33.9\%$ for IncrementDET and retrained SpecTTTra. When the thresholds are instead set on one half of the degraded human test tracks and evaluated on the other half, the false-positive rate returns to $4.5$--$5.4\%$, and IncrementDET keeps a higher TPR than SpecTTTra under every update strategy. Against retraining, it reaches $93.2\%$ against $49.4\%$ under MP3 encoding and $84.1\%$ against $44.8\%$ under resampling. The calibration set should therefore contain the processing that is expected in deployment. For IncrementDET, recalibration needs only human tracks with the same processing and no retraining.

\begin{table}[h]
\centering
\small
\setlength{\tabcolsep}{3pt}
\caption{Robustness of the models of the last step of the continual comparison. Top: EER, mean over the twelve generators. Bottom: false-positive rate on human test tracks at the $5\%$ target, with thresholds set on the clean calibration set. All values in percent.}
\label{tab:degrade_full}
\begin{tabular}{@{}lcccccc@{}}
\toprule
& & \multicolumn{2}{c}{MP3 (kbit/s)} & Resample & \multicolumn{2}{c}{Shelf (dB)} \\
\cmidrule(lr){3-4}\cmidrule(l){6-7}
& Clean & 128 & 64 & 11\,kHz & $+6$ & $-6$ \\
\midrule
\multicolumn{7}{@{}l}{\emph{EER}} \\
IncrementDET & \textbf{1.9}  & \textbf{9.2}  & \textbf{9.2}  & \textbf{14.3} & \textbf{2.1}  & \textbf{2.0}  \\
Retrain      & 9.3  & 21.5 & 21.5 & 24.5 & 9.8  & 9.7  \\
Replay       & 13.5 & 26.0 & 25.8 & 29.6 & 14.1 & 14.1 \\
EWC          & 29.7 & 30.4 & 30.4 & 37.3 & 30.8 & 30.3 \\
Fine-tune    & 35.5 & 38.6 & 38.6 & 38.7 & 36.3 & 35.3 \\
No update    & 31.0 & 37.9 & 37.8 & 36.2 & 32.2 & 32.5 \\
\midrule
\multicolumn{7}{@{}l}{\emph{FPR on human test tracks, clean thresholds}} \\
IncrementDET & 3.5 & 33.9 & 33.3 & 30.6 & 4.7 & 4.0 \\
Retrain      & 3.8 & 26.4 & 25.8 & 25.6 & 4.2 & 3.5 \\
Replay       & 4.4 & 18.7 & 18.4 & 50.0 & 5.6 & 3.8 \\
EWC          & 5.1 & 0.4  & 0.4  & 37.8 & 6.0 & 4.2 \\
Fine-tune    & 4.9 & 0.6  & 0.6  & 46.8 & 6.4 & 4.2 \\
No update    & 4.7 & 10.4 & 10.6 & 27.1 & 7.4 & 4.8 \\
\bottomrule
\end{tabular}
\end{table}

\begin{table}[h]
\centering
\small
\setlength{\tabcolsep}{3.5pt}
\caption{Robustness of the models of the last step of the continual comparison. Mean TPR at $5\%$ FPR over the twelve generators, with thresholds set on degraded human tracks. MP3 at $128$\,kbit/s. All values in percent.}
\label{tab:degrade}
\begin{tabular}{@{}lccccc@{}}
\toprule
& Clean & MP3 & Resample & Low $+6$ & High $-6$ \\
\midrule
IncrementDET & \textbf{96.6} & \textbf{93.2} & \textbf{84.1} & \textbf{96.6} & \textbf{96.8} \\
Retrain      & 84.1 & 49.4 & 44.8 & 83.5 & 82.8 \\
Replay       & 74.2 & 34.0 & 32.1 & 70.5 & 72.4 \\
EWC          & 28.8 & 26.4 & 18.3 & 28.2 & 28.4 \\
Fine-tune    & 28.7 & 23.3 & 22.7 & 28.0 & 29.0 \\
No update    & 30.5 & 14.0 & 18.8 & 27.6 & 27.4 \\
\bottomrule
\end{tabular}
\end{table}

\section{Limitations and future work}
\label{sec:supp_limits}
This section extends the discussion of Sec.~\ref{sec:discussion}. 

\subsection{Preprocessing trade-offs}
In our benchmark, all audio is resampled to $16$\,kHz mono and loudness-normalized, which removes cues from how platforms store audio (Sec.~\ref{sec:dataset}). This preprocessing also removes the content above $8$\,kHz, so our benchmark is not suited to forensic methods that analyze full-bandwidth residuals. We accept this cost because cues that disappear after a single resampling operation are not reliable detection signals. Future work can add a full-bandwidth track to the benchmark, in which all sources are stored in one common format.

\subsection{Model updates and private models}
For IncrementDET, a needle describes a generator as sampled through a particular product version. If a commercial service changes its model without notice, the needle may become outdated, and the generator must be enrolled again with new samples. We have not measured how quickly a needle becomes outdated. The question for future work is whether a detector can notice automatically that the output of an enrolled generator has changed and decide when the generator should be enrolled again. Because one enrollment takes seconds, the new enrollment itself is inexpensive once the change is noticed. Furthermore, private models are covered only through transfer. IncrementDET enrolls a generator from samples of its output. Privately fine-tuned models that are never released cannot be enrolled until some of their outputs are identified, and until then they are detected only to the extent that the existing needles transfer to them. The family transfer of \cref{sec:supp_grouping} suggests that the needle of an enrolled generator may transfer to fine-tuned variants of that generator, which future work can test.

\subsection{Maintenance of the human data}
Every detector that relies on human reference data depends on a dataset of human-made music. Such a dataset ages and, as \cref{sec:supp_data} shows, can contain generated uploads from open platforms. Datasets recent enough to match the style of current music are also the ones most exposed to recent generated uploads. Maintaining the human data is therefore a recurring task with a measurable effect at strict thresholds. The audit of \cref{sec:supp_data} shows that the needles themselves can flag generated uploads in the human data, and automating this screening is future work.

\subsection{Potential harm}
A false positive labels the work of a human musician as generated, and a platform that acts on such a label can affect the income and reputation of that musician. A platform should therefore operate the detector at a low false-positive rate, which is why we report the TPR at false-positive rates of $5\%$ and $1\%$ (Sec.~\ref{sec:metrics}). IncrementDET lets a platform set the overall false-positive target $\alpha$ directly, and at the $1\%$ target it still detects $90.8\%$ of the tracks of the twelve enrolled generators. A platform should calibrate the detector on its own catalog, and it should combine a positive output with further evidence before it acts against a musician.

\end{document}